\documentclass[aps,prl,amsmath,amssymb,reprint,superscriptaddress,longbibliography]{revtex4-1}

\pdfoutput=1
\usepackage{graphicx}  % needed for figures
\usepackage{bm}        % for math
\usepackage{amssymb}   % for math
\usepackage{lettrine}
\usepackage{soul}
\usepackage{color}
\usepackage{braket}
\usepackage{tabularx}
\usepackage{subcaption}
\begin{document}

\title{Ultra-High Cooperativity Interactions between Magnons and Resonant Photons in a YIG sphere}
%Ultra-High Cooperativity Coupling between dielectric photon resonances and magnon spin wave resonaces in a YIG sphere}
	
\author{J. Bourhill}
\email{jeremy.bourhill@uwa.edu.au}
\author{N. Kostylev}
\author{M. Goryachev}
\author{D. L. Creedon}
\author{M.E. Tobar}	

\affiliation{ARC Centre of Excellence for Engineered Quantum Systems, University of Western Australia, 35 Stirling Highway, Crawley WA 6009, Australia}

\date{\today}

% the following line is for submission, including submission to the arXiv!!
%\hspace{5.2in} \mbox{Fermilab-Pub-04/xxx-E}

\begin{abstract}
\noindent \textit{Resonant photon modes of a 5mm diameter YIG sphere loaded in a cylindrical cavity in the 10-30GHz frequency range are characterised as a function of applied DC magnetic field at millikelvin temperatures. The photon modes are confined mainly to the sphere, and exhibited large mode filling factors in comparison to previous experiments, allowing ultrastrong coupling with the magnon spin wave resonances. The largest observed coupling between photons and magnons is $2g/2\pi=7.11$ GHz for a 15.5 GHz mode, corresponding to a cooperativity of $C=1.51\pm0.47\times10^7$. 
Complex modifications beyond a simple multi-oscillator model, of the photon mode frequencies were observed between 0 and 0.1 Tesla. Between 0.4 to 1 Tesla, degenerate resonant photon modes were observed to interact with magnon spin wave resonances with different couplings strengths, indicating time reversal symmetry breaking due to the gyrotropic permeability of YIG. Bare dielectric resonator mode frequencies were determined by detuning magnon modes to significantly higher frequencies with strong magnetic fields.  By comparing measured mode frequencies at 7 Tesla with Finite Element modelling, a bare dielectric permittivity of $15.96\pm0.02$ of the YIG crystal has been determined at about $20 mK$.
}
\end{abstract}

\maketitle

\section{Introduction}

Hybrid photon-magnon systems in ferromagnetic spheres have recently emerged as a promising approach towards coherent information processing \cite{Chumak,Tabuchi2,Tabuchi,Zhang,Goryachev,Bai,Huebl,Zhang4}. Due to the large exchange interaction between spins in ferromagnets, they will lock together to form a macrospin that can be utilised for coherent information processing protocols \cite{PhysRevLett.102.083602,Zhang4}. The quantised excitation of the collective spin is referred to as a magnon. Yttrium Iron Garnet (YIG) based magnon systems are attractive due to very high spin density, resulting in significant cooperativity as well as relatively narrow linewidths \cite{Goryachev,Zhang2,Osada,Zhang}. Furthermore, due to the possibility of coupling magnon modes to photons at optical frequencies \cite{Zhang2,Osada,Shen,Demokritov}, magnon systems may be considered as a candidate for coherent conversion of microwave and optical photons \cite{Zhang2,Osada}. In addition, magnons interact with elastic waves \cite{Kittel,Zhang3} opening a window for combining mechanical and magnetic systems. These systems therefore possesses great potential as an information transducer that mediates inter-conversion between information carriers of different physical nature thus establishing a novel approach to hybrid quantum systems \cite{PhysRevLett.92.247902,PhysRevLett.102.083602,PhysRevLett.103.043603,Xiang:2013aa}. 
%This field has been termed {``}optomagnonics{''} due to the syst em{'}s similarities to optomechanics \cite{Zhang2,Osada}. 

% The magnon{'}s great tunability with DC magnetic field and long lifetime \cite{Goryachev,Zhang2,Osada,Zhang} make it an ideal information carrier. It can couple to microwaves through a magnetic dipole interaction, as well as elastic waves \cite{Kittel,Zhang3} and optical light \cite{Zhang2,Osada,Shen,Demokritov}. I

Among all magnon systems the central role is devoted to YIG, a material that possesses exceptional magnetic and microwave properties and has been used in microwave systems such as tuneable oscillators and filters for many decades \cite{lvov,gurevich}. Although, only recently 
Soykal and Flatt\'e proposed and modelled the photon-magnon interaction based on YIG nano-spheres with application to quantum systems \cite{Soykal,Soykal2}. As predicted by the authors, extremely large coupling rates, $g$, could be achieved in YIG spheres, which is favourable for coherent information exchange and has been demonstrated experimentally later \cite{Goryachev,Kostylev,Zhang}. For these experiments, the interaction is observed between photon and magnon resonances created correspondingly by photon cavity boundary conditions and spin precession under external DC magnetic field. A commonly used method is to place a relatively small YIG sphere in a local maxima of the magnetic field inside a much larger microwave cavity. This is done to achieve quasi uniform distribution of the cavity field over the sphere volume to avoid spurious magnon modes. Cavities can take oval \cite{Tabuchi,Zhang} or spherical shapes \cite{Rameshti,Soykal,Soykal2}, and even re-entrant cavities with multiple posts have been used in an attempt to focus the microwave energy over the sphere \cite{Goryachev,Kostylev}. In this work we investigate a completely different regime in which the magnon and photon wavelengths are comparable, leading to considerably larger coupling strengths, but additional couplings to higher order modes. 
In general, for this case the strength of the photon-magnon interaction will be determined by an overlap integral of the two respective mode shape functions. Given the magnon mode shape is limited to the sphere{'}s volume, this integral will be maximised when the photon mode is confined to the same volume. To achieve the latter, we utilise an exceptionally large YIG sphere with diameter $d=5$ mm, matching magnon and microwave photon mode volumes, unlike previous microwave cavity experiments.
%Due to this large size, a majority of the resonant microwave field resides inside the YIG (unlike previous microwave cavity experiments), hence maximising the overlap of photonic modes with the magnon resonances. 

%In general, the strength of the photon-magnon interaction will be determined by an overlap integral of the two respective mode shape functions. 

% The regime where photon and magnon mode wave-numbers are comparable. 

%Thus, concerning the microwave domain it is important to understand the photon-magnon interaction in a regime that has not yet been investigated, i.e.

%In this work, the resonant microwave modes are described in \cite{spheres}.

In order to investigate this regime we use common microwave spectroscopy techniques \cite{Farr,gyro2,gyro3,Bourhill1,Goryachev,Kostylev} to directly observe the mode splitting caused by the magnon interaction to determine the coupling values. Similar systems have been extensively utilised not only in the field of spintronics to investigate the interaction between microwave photons and paramagnetic spin ensembles \cite{Farr,Bourhill1,gyro2,gyro3,ruby}, but also to realise optical comb generation \cite{delhaye}, ultra-low threshold lasing \cite{Kippenberg}, cavity-assisted cooling, control and measurement of optomechanical systems \cite{Aspelmeyer,Bourhill2}, and extremely stable cryogenic sapphire oscillator clock technology \cite{ivanov3,ivanov4}.
To date, exciting internal, highly-confined photonic modes in a YIG sphere has only recently been demonstrated in the optical regime using Whispering Gallery Modes (WGM) \cite{Zhang2,Osada} but has never before been achieved in the microwave domain. This is due to the typical sub-millimetre diameter of the spheres. As such, interactions with magnons must be observed via Brillouin scattering \cite{PhysRevB.86.134403}, which has yielded high quality factors and also demonstrated a pronounced nonreciprocity and asymmetry in the sideband signals generated by the magnon-induced scattering.

Extremely large mode splittings ($g/\omega>0.1$) cause simultaneous coupling to a higher density of modes, with an overlap of avoided level crossings. Therefore, the model proposed by Soykal and Flatt\'e \cite{Soykal} becomes no longer applicable, as it assumes the interaction occurs between a single photonic and magnon mode.
More recently, a paper by Rameshti \textit{et al.} \cite{Rameshti} simulated a similar scenario of the presented experiment, in which the ferromagnetic sphere is itself the microwave cavity. Our observed results may appear to be in good agreement with this work{'}s predictions, however, what is apparent is that in this specialised case, one must consider more than just the magnetostatic, uniform Kittel magnon mode, a limitation of \cite{Rameshti}.
Indeed, due to the nonuniformity of both the microwave mode magnetic field energy density across the sphere, which is unique to this experiment, and the nonuniformity of the sphere parameters arising due to cryogenic cooling, the assumption that only the uniform Kittel magnon resonance participates is no longer valid. Despite this, in this paper we use a two mode model to obtain estimations of coupling strengths, and demonstrate how this results in inconsistent susceptibility values.

\begin{figure}[t!]
\centering
\includegraphics[width=0.5\textwidth]{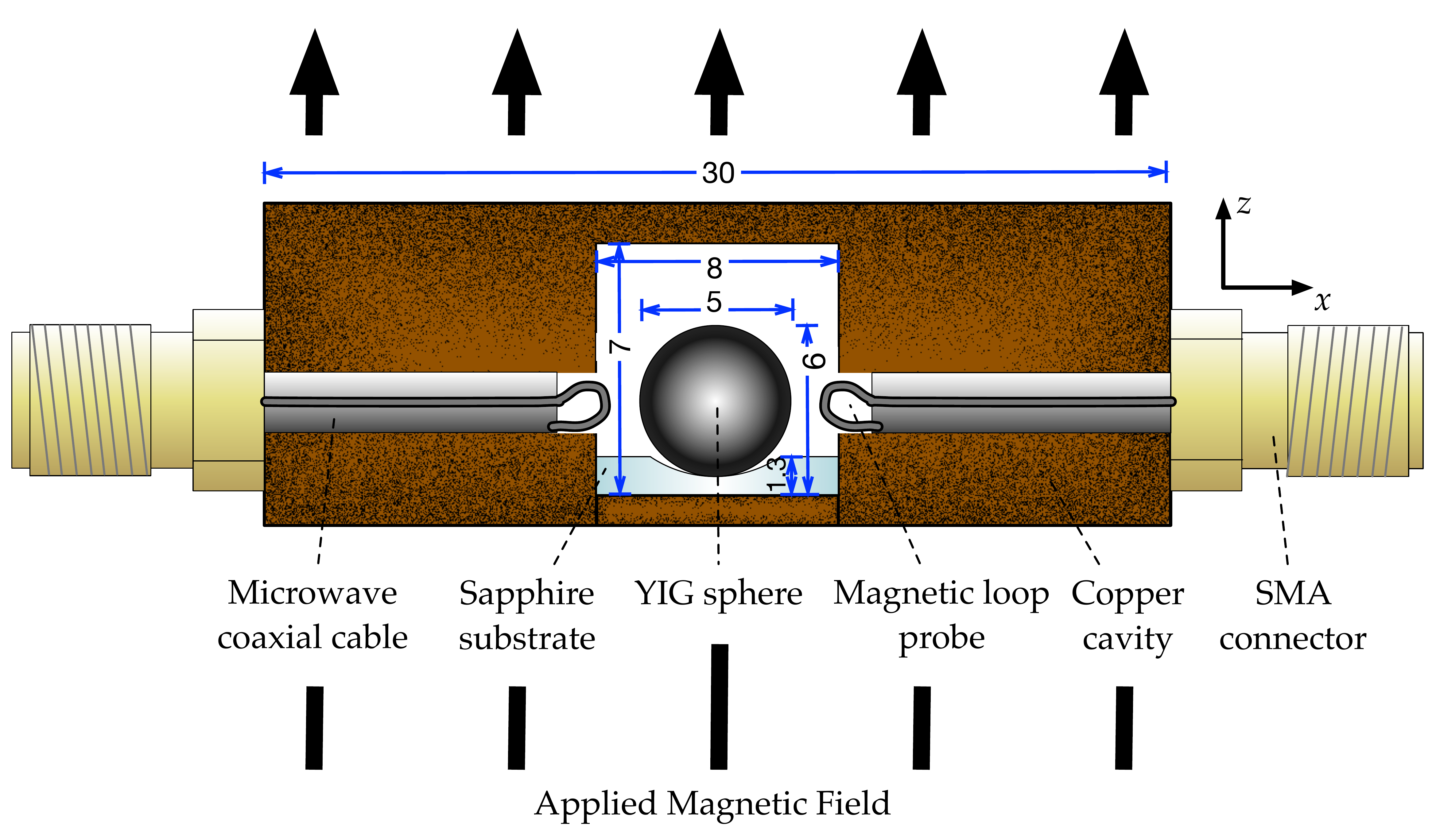}
\caption{\small{(Color online) Cross section of the copper cavity that houses the 5 mm YIG sphere. The sphere sits on a sapphire disk, and microwaves are coupled in and out of it via loop probes which produce and detect an $H_\phi$ component. A variable DC magnetic field is applied along the $z$ axis.}}
\label{fig:setup}
\end{figure}

\section{Physical Realization}
The $d=5$ mm YIG sphere was manufactured by Ferrisphere, Inc. with a quoted room temperature  saturation magnetisation of $\mu_0 M=0.178$ T. It is placed on a small sapphire disk, with a concavity etched out using a diamond tipped ball grinder, to keep the sphere from rolling out of position, and reduce dielectric losses that would arise if the YIG were in direct contact with the conductive copper housing. Sapphire was chosen over teflon as an intermediary between the YIG and copper to improve the thermal conductivity to the sphere. 

Together, the sapphire and YIG are housed inside a copper cavity with dimensions specified in figure \ref{fig:setup}. A loop probe constructed from flexible subminiature version A cable launchers is used to input microwaves and a second is used to make measurements, allowing the determination of $S$ parameters. The entire cavity is cooled to about 20 mK by means of a dilution refrigerator (DR) with a cooling power of about 500 $\mu$W at 100 mK. The cavity is attached to a copper rod bolted to the mixing chamber stage of the DR that places it at the field center of a 7 T superconducting magnet, whose applied field is oriented in the $z$ direction of the cavity. The magnet is attached to the 4 K stage of the DR, with the copper cavity mounted within a radiation shield of approximately 100 mK that sits within the bore of the magnet.

\section{Experimental Observations}
	
	 \begin{figure}[t!]
    \centering
    \includegraphics[width=0.5\textwidth]{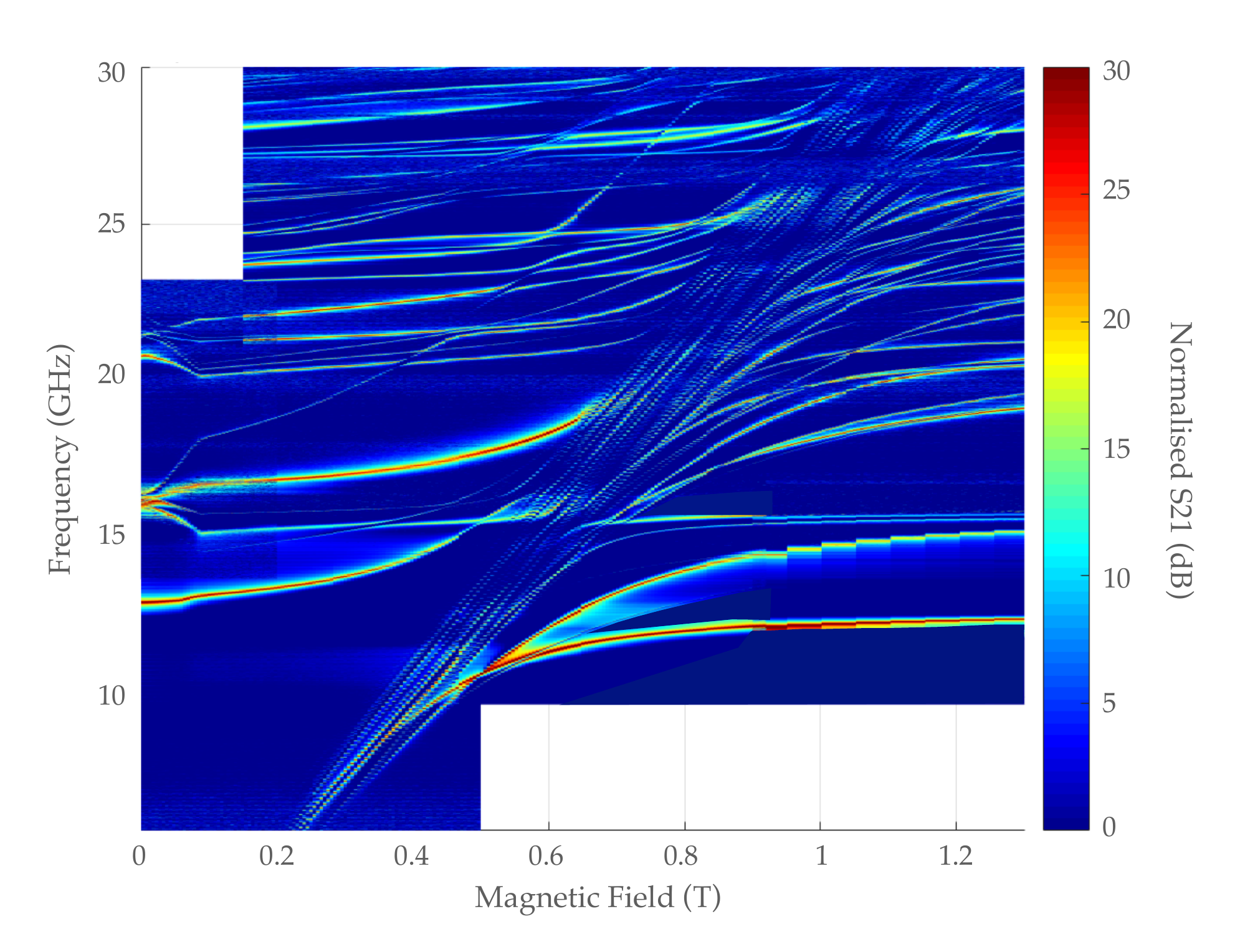}
    \caption{\small{(Color online) Transmission data as magnetic field is swept.}}
    \label{fig:full_spec}
    \end{figure}
The transmission spectrum of the YIG was recorded for DC magnetic fields swept from 0 {--} 7 T using a vector network analyser (VNA), with partial results shown in figure \ref{fig:full_spec}. A host of magnon resonances/higher order magnon-polaritons can be observed originating from (0 T, 0 GHz) with an approximate gradient of 28 GHz/T. The more-or-less horizontal lines approaching the magnon resonances from either side correspond to resonant photon modes of the sphere. Importantly, we can observe that in the dispersive regime, far removed from any microwave resonant mode, there still exist multiple magnon modes. We observe that the anticrossing gaps are populated by unperturbed modes, which are remnant {``}tails{''} of both {``}higher{''} and {``}lower{''} mode interactions, as predicted by Rameshti \textit{et al.} in the ultrastrong coupling regime \cite{Rameshti}. 

\begin{figure}[b!]
    \centering
    \includegraphics[width=0.5\textwidth]{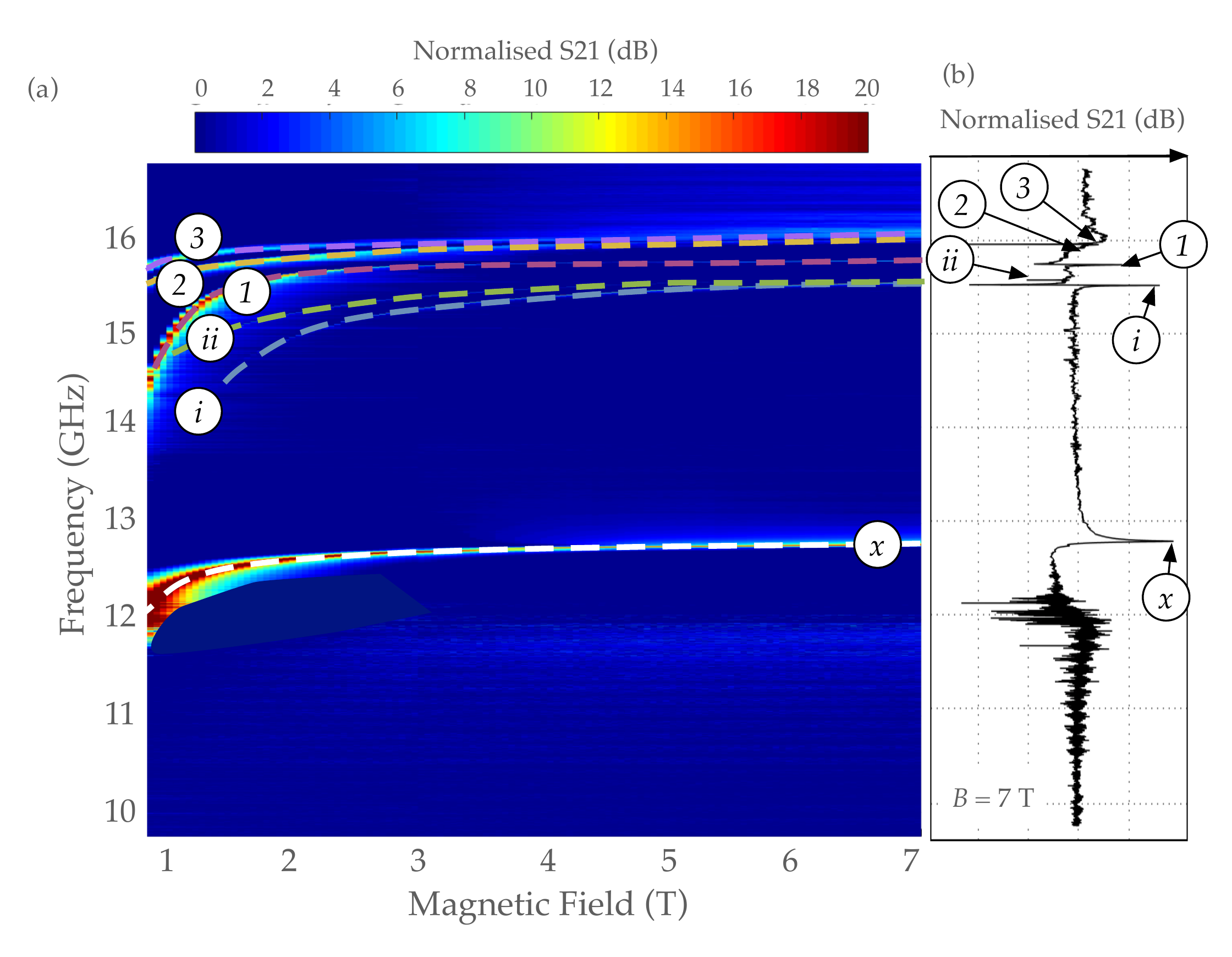}
    \caption{\small{(Color online) (a) Asymptotic frequency values of the six lowest order photon modes as $B\to 7$ T. (b) Transmission spectra at $B=7$ T, from which mode linewidths may be measured.}}
    \label{fig:highfield}
    \end{figure}

For the remainder of this paper, we will focus on the six lowest frequency photon modes, whose resonant frequencies may only be accurately determined at large magnetic fields, when the entire spin ensemble has been detuned, as shown by figure \ref{fig:highfield}.
    
 The modes have been categorised into three distinct classes: mode {``}$x${''} is the lowest frequency and lowest $Q$ factor mode, the two highest $Q$ factor modes; {``}$i${''} and {``}$ii${''}, and the three remaining highest frequency modes; {``}1{''}, {``}2{''} and {``}3{''}. Their asymptotic frequencies as $B\to7$ T are summarised in table \ref{tab:freqs}.
 
\begin{figure}[t!]
    \centering
    \includegraphics[width=0.45\textwidth]{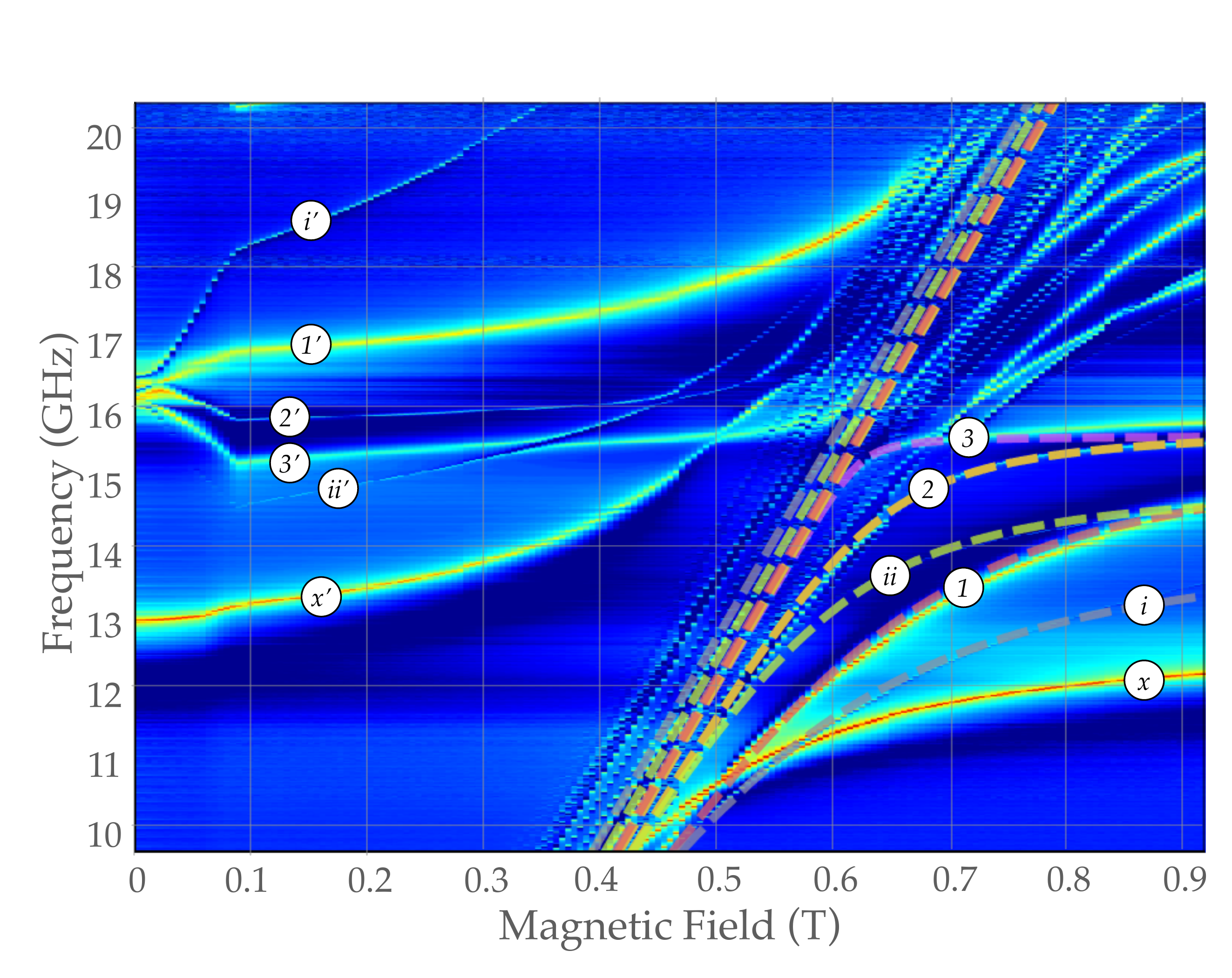}
    \caption{\small{(Color online) Two harmonic oscillator model fitting to modes $i$, $ii$, 1, 2 and 3. From the curved lineshapes, one can determine the coupling value $g$.}}
    \label{fig:modefits}
    \end{figure}

\begin{figure}[h!]
\centering
\includegraphics[width=0.5\textwidth]{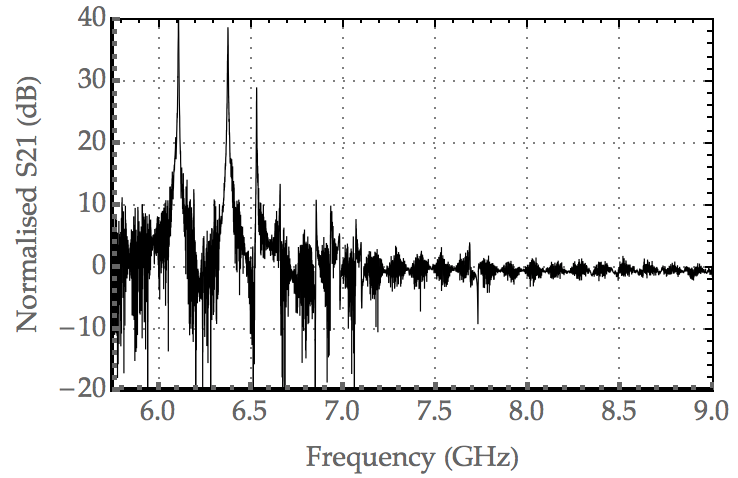}
\caption{S21 transmission spectra showing a host of magnon resonance peaks at $B=0.2475$ T.}
\label{fig:magnons}
\end{figure}

\begin{figure}[h!]
\centering
\includegraphics[width=0.5\textwidth]{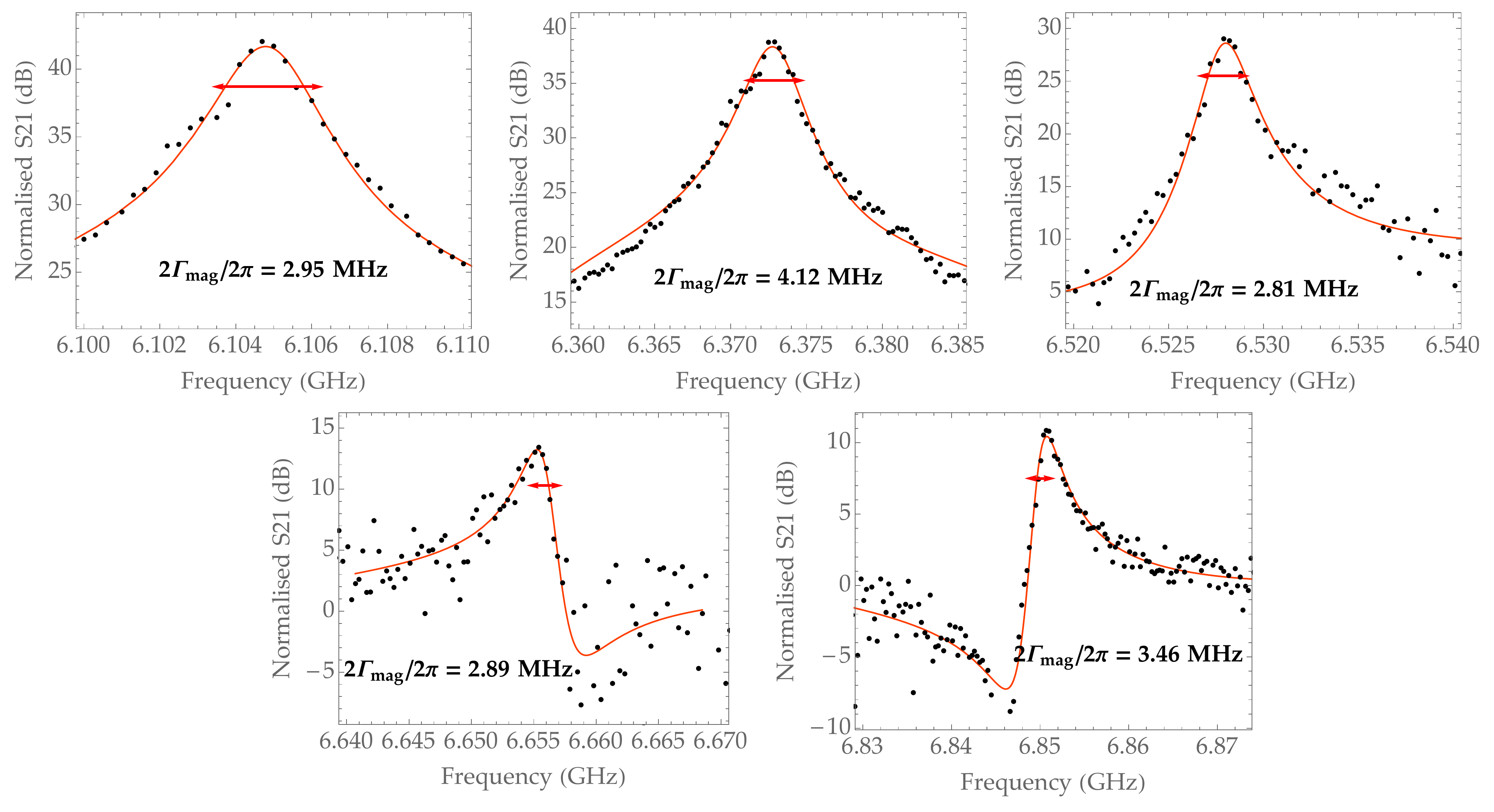}
\caption{Fitting the magnon resonances with Fano fits.}
\label{fig:magnonfits}
\end{figure}

The behaviour of these modes as the magnon resonances are tuned via the applied magnetic field is shown in figure \ref{fig:modefits}. It has been shown previously \cite{Goryachev} that a standard model of two interacting harmonic oscillators can accurately determine the coupling values from such avoided crossings. However, we observe strong distortion around 0 T, and also an asymmetry of the mode splittings about the central magnon resonances due to the ultrastrong coupling of the photon modes to the magnon modes, as was observed previously in Ruby \cite{ruby}. Therefore we fit only the curves to the right of the magnon resonance. These fits are shown as the dashed lines in figure \ref{fig:modefits}. From these fits we can approximate the values of $g$ for each mode, as summarised in table \ref{tab:freqs}. The linewidths, $\Gamma_j$ and frequencies, $\omega_j/2\pi$ of the photon modes are determined from the transmission spectra taken at high field values (figure \ref{fig:highfield} (b)), whilst the magnon linewidth, $\Gamma_\text{mag}$ can be determined by analysing the transmission spectra in the dispersive regime. We take a frequency sweep at $B=0.2475$ T from 5.75{--}9 GHz in order to view the magnon resonance peaks far away from any interaction with the dielectric microwave modes, as shown in figure \ref{fig:magnons}. There is a level of variation amongst the magnon linewidths as calculated by fitting the peaks with Fano resonance fits, as shown in figure \ref{fig:magnonfits}. This variation and the presence of multiple peaks demonstrates the presence of higher order magnon modes. Taking the average and standard deviation of these linewidths gives a final estimate of magnon linewidth as $\Gamma_\text{mag}/\pi=3.247 \pm 0.493 $ MHz. Cooperativity is calculated as $C_j=g_j^2/\Gamma_\text{mag}\Gamma_j$.
%Due to the ultrastrong coupling however, it is difficult to say definitively if these modes are pure magnetic, or if there is still some level of hybridisation with the dielectric modes. Indeed, 
      \begin{table}[t!]
        \centering
        \begin{tabular}{|c|c|c|c|c|c|}
        \hline 
        {~}Mode{~} & {~}$\omega_{j\left| B\to\text{7 T}\right.}/2\pi${~} & {~}$\Gamma_j/\pi${~} & {~}$g_j/\pi${~} & $C_j$&$g_j/\omega_j$\\
        &(GHz)&(MHz)&(GHz)&($\times10^5$)&(\%)\\ \hline
        %&&&&\\ 
        $x$ & 12.779 & 11.84 & 4.79 & 5.97$\pm$1.85&18.7\\ \hline
        %&&&&$\times10^{5}$\\ \hline
        %&&&&\\ 
        $i$ & 15.506 & 1.029 & 7.11 &151$\pm$47.0&22.9\\ \hline 
        %&&&&$\times10^{7}$\\ \hline
        %&&&&\\ 
        $ii$ & 15.563 & 1.197 & 4.19 & 45.2$\pm$14.0&13.5\\ \hline
        %&&&&$\times10^{6}$\\ \hline
        %&&&&\\ 
        1 & 15.732 & 5.355 & 6.15 & 21.8$\pm$6.76&19.5\\  \hline
        %&&&&$\times10^{6}$\\ \hline
        %&&&&\\ 
        2 & 15.893 & 2.965 & 3.04 & 9.60$\pm$2.98&9.56\\ \hline
        %&&&&$\times10^{5}$\\ \hline
        %&&&&\\ 
        3 & 15.950 & 2.965 & 0.78 & 0.632$\pm$0.196&2.45\\ \hline
        %&&&&$\times10^{4}$\\ \hline
        \end{tabular}
        \caption{Measured and calculated results for each mode showing couplings $g_j$ and cooperativities, $C_j$.}
        \label{tab:freqs}
        \end{table}
    
The cooperativity values in table \ref{tab:freqs} demonstrate that all modes are strongly coupled to the magnons, and all with the exception of modes 2 and 3 are in the ultrastrong coupling regime (i.e. $g_j/\omega_j\geq 0.1$ \cite{Rameshti}). The largest cooperativity value obtained is that of mode $i$, which is, to the authors{'} knowledge, the largest value ever reported to date in any previously studied spin system. 

\begin{figure}[t!]
    \centering
    \includegraphics[width=0.5\textwidth]{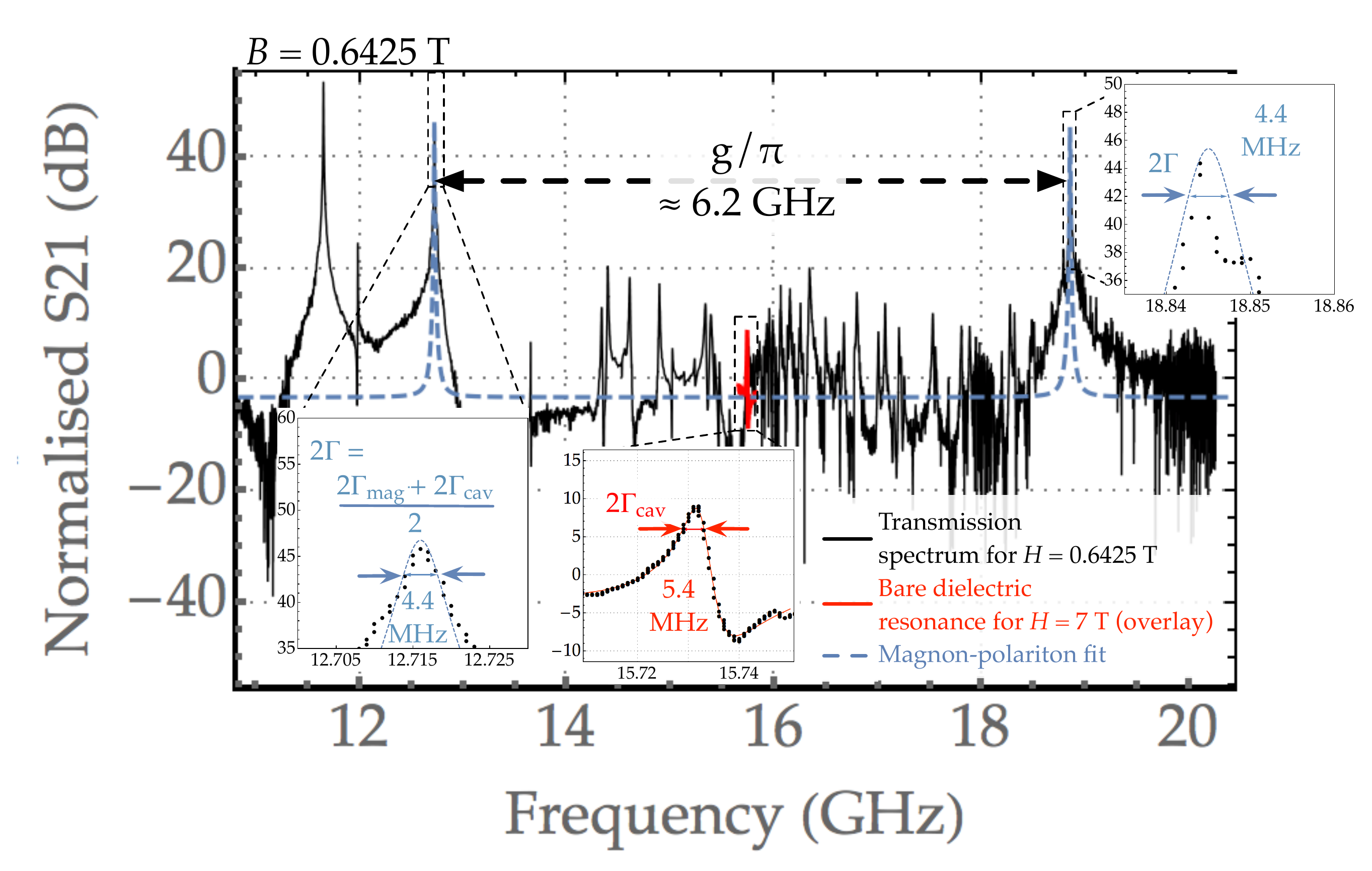}
    \caption{\small{(Color online) Transmission spectrum at $B=0.6425$ T. At this applied magnetic field the magnon resonance is tuned coincident in frequency with mode 1, and the strong coupling between the two results in a mode splitting of 6.2 GHz. The high density of resonant peaks in the centre of the figure suggest a large number of higher order magnon modes are present in this system.}}
    \label{fig:modesplitting}
    \end{figure}

A transmission spectrum taken at $B=0.6425$ T is shown in figure \ref{fig:modesplitting}, demonstrating the mode splitting of mode 1, symmetric about the magnon resonance. Overlaid in red is the bare photon resonance at 7 T, i.e. the microwave mode unperturbed by the magnon modes. From this red curve, $2\Gamma_1/2\pi$ is determined to be 5.355 MHz, as shown in table \ref{tab:freqs}. When one takes the average of $2\Gamma_\text{mag}/2\pi=3.247$ MHz and $2\Gamma_1/2\pi$, one obtains the line width of the resulting hybrid state when the magnon resonance is tuned coincident in frequency with the photon mode, as depicted by the dashed blue curve in figure \ref{fig:modesplitting}, i.e., $\sim$ 4.4 MHz. This excellent agreement indicates that at this particular $B$ field, mode 1 exists as a hybrid magnon-polariton.
    
    \begin{figure}[b!]
    \centering
    \includegraphics[width=0.5\textwidth]{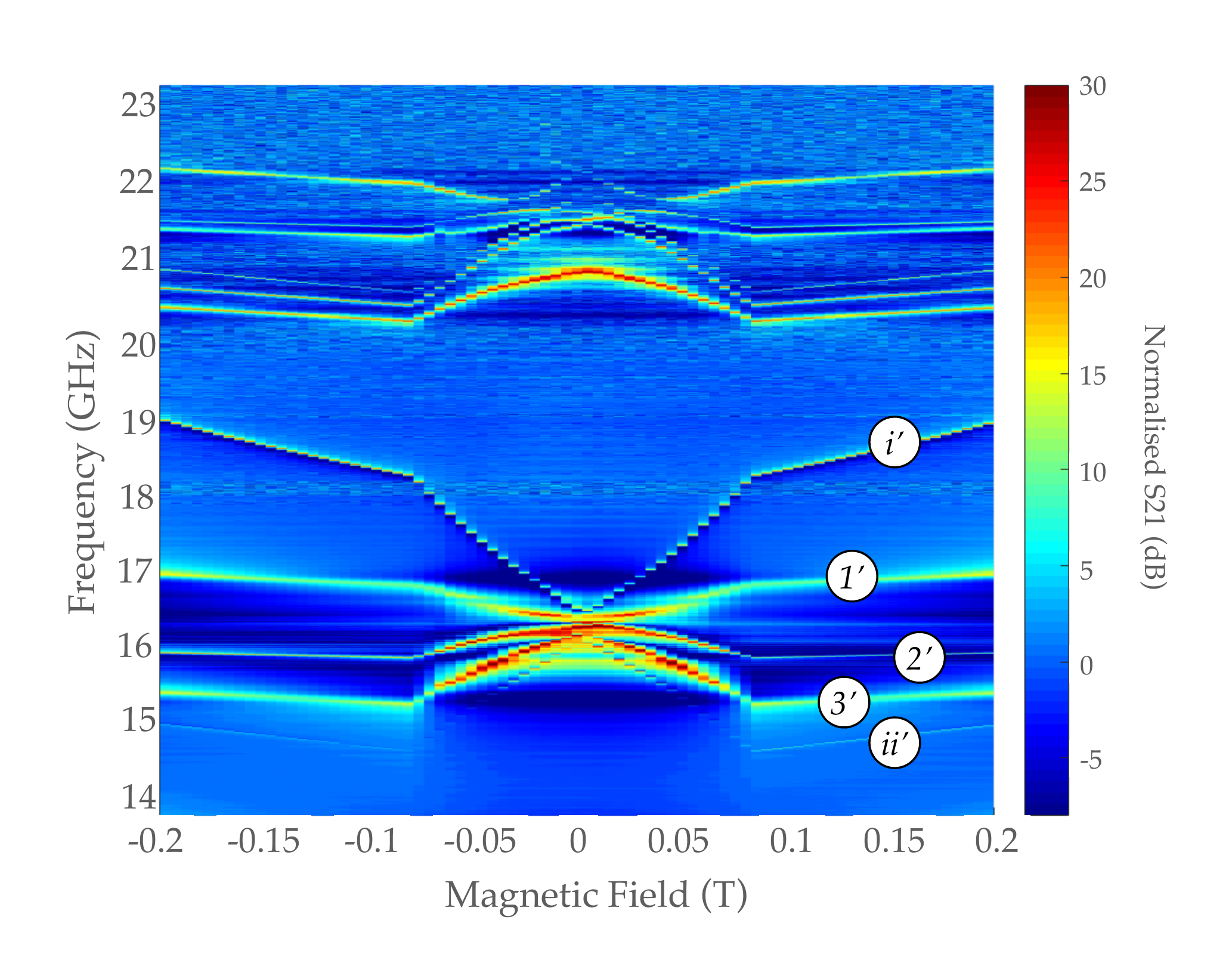}
    \caption{\small{(Color online) Behaviour of the photon modes around $B=0$ T, a result of the internal magnetisation of YIG.}}
    \label{fig:DC}
    \end{figure}
    
Around $B=0$ T, we observe a severe distortion of the cavity mode frequency dependence on magnetic field as demonstrated by Fig.~\ref{fig:DC}. Around 16 GHz, we see there exist five modes, corresponding to modes $i$, $ii$ and 1{--}3, on the {``}left{''} side of the magnon resonances. These modes have been given a primed nomenclature to indicate their existence at $B$ fields lower than that required to tune the magnons to their frequencies. This phenomenon has been previously observed in single crystal YAG \cite{YAG} highly doped with rare-earth Erbium ions, and is explained by the influence on the ferromagnetic phase of the impurity ions on degenerate modes. The effect can be explained by the influence of the ensemble of strongly coupled spins on the centre-propagating waves of the near degenerate mode doublet. For large spin-photon interactions, tails of Avoided Level Crossings (ALCs) from the positive half plane ($B>0$) should still exist on the negative half plane ($B<0$) and vice a versa. Although, instead of gradual change of direction, the system demonstrates an abrupt transition to a {``}no coupling{''} state. It is worth mentioning that such effect has not been observed in photonic systems interacting with paramagnetic spin ensembles \cite{Farr,gyro2,PhysRevA.89.013810}. In the present case, the effect is much more pronounced, with fractional frequency deviations and the magnetic field range of the effect both orders of magnitude larger than observed previously \cite{YAG}, a result of the magnetic spin density. %This is because the Erbium ions in YAG behave as a {``}dilute{''} ferromagnet compared to YIG.

%But each of these ALC is associated with magnons of particular direction of magnetisation that do not exist on the other half plane. In such a way that a the positive ALC tail can exist only on the positive half plane and must to be interrupted at some point.

%That means that on both sides of the plane, photons interact with both positively magnetised ma

% The same logic should apply to the nonuniform magnon modes interacting with the WGM doublet and following the same field patterns and changing a corresponding resonance doublet at zero field. On the other hand that implies that at near zero field each magnon modes is coupled to each of two WGM doublet resonance.

%The convergence of mode frequencies occurs at a magnetic field at which tails from interactions in the positive half plane ($B>0$) and negative half plane ($B<0$) meet. Without this feature photons would interact simultaneously with {``}spin up{''} transitions from $B>0$ and {``}spin down{''} from $B<0$. This is permissible in paramagnetic materials where all spins are not correlated so both types can exist and the positive tails gradually decay into negative $B$ values. In ferromagnets, this is not possible since all spins are coherent, so the modes collapse immediately around 0 T to preserve the coherence inside the sample. 

\section{Discussion}
COMSOL 3.5{'}s electromagnetic package was used to model the system. A 3D model was used so as to analyse the degeneracies in the $\phi$ axis of the dielectric modes. The internal copper wall of the cavity is modelled as a perfect electrical conductor, which for the purposes of the desired eigenfrequency study, is an appropriate simplification. 
%As COMSOL will only give solutions in cartesian coordinates, it was necessary to convert the electric and magnetic fields into spherical coordinates according to
%	\begin{equation}
%	\begin{array}{ll}
%	\left\{\begin{matrix}
%	E_\rho \\
%	H_\rho
%	\end{matrix}\right\}=&
%	\left\{\begin{matrix}
%	E_x \\
%	H_x
%	\end{matrix}\right\}\sin\left(\arccos\left(\frac{z}{\sqrt{x^2+y^2+z^2}}\right)\right)\cos\left(\arctan\left(\frac{y}{x}\right)\right)\\
%	&\\
%	&+\left\{\begin{matrix}
%	E_y \\
%	H_y
%	\end{matrix}\right\}\sin\left(\arccos\left(\frac{z}{\sqrt{x^2+y^2+z^2}}\right)\right)\sin\left(\arctan\left(\frac{y}{x}\right)\right)\\
%	&\\
%	&+\left\{\begin{matrix}
%	E_z \\
%	H_z
%	\end{matrix}\right\}\left(\frac{z}{\sqrt{x^2+y^2+z^2}}\right),\\
%	&\\
%	\left\{\begin{matrix}
%	E_\phi \\
%	H_\phi
%	\end{matrix}\right\}=&
%	\left\{\begin{matrix}
%	E_y \\
%	H_y
%	\end{matrix}\right\}\cos\left(\arctan\left(\frac{y}{x}\right)\right)-
%	\left\{\begin{matrix}
%	E_x \\
%	H_x
%	\end{matrix}\right\}\sin\left(\arctan\left(\frac{y}{x}\right)\right),\\
%	&\\
%	\left\{\begin{matrix}
%	E_\theta \\
%	H_\theta
%	\end{matrix}\right\}=&
%	\left\{\begin{matrix}
%	E_x \\
%	H_x
%	\end{matrix}\right\}\left(\frac{z}{\sqrt{x^2+y^2+z^2}}\right)\cos\left(\arctan\left(\frac{y}{x}\right)\right)\\
%	&\\
%	&+\left\{\begin{matrix}
%	E_y \\
%	H_y
%	\end{matrix}\right\}\left(\frac{z}{\sqrt{x^2+y^2+z^2}}\right)\sin\left(\arctan\left(\frac{y}{x}\right)\right)\\
%	&\\
%	&+\left\{\begin{matrix}
%	E_z \\
%	H_z
%	\end{matrix}\right\}\sin\left(\arccos\left(\frac{z}{\sqrt{x^2+y^2+z^2}}\right)\right).
%	\end{array}
%	\end{equation}

\begin{figure*}
\centering
\includegraphics[width=0.6\textwidth]{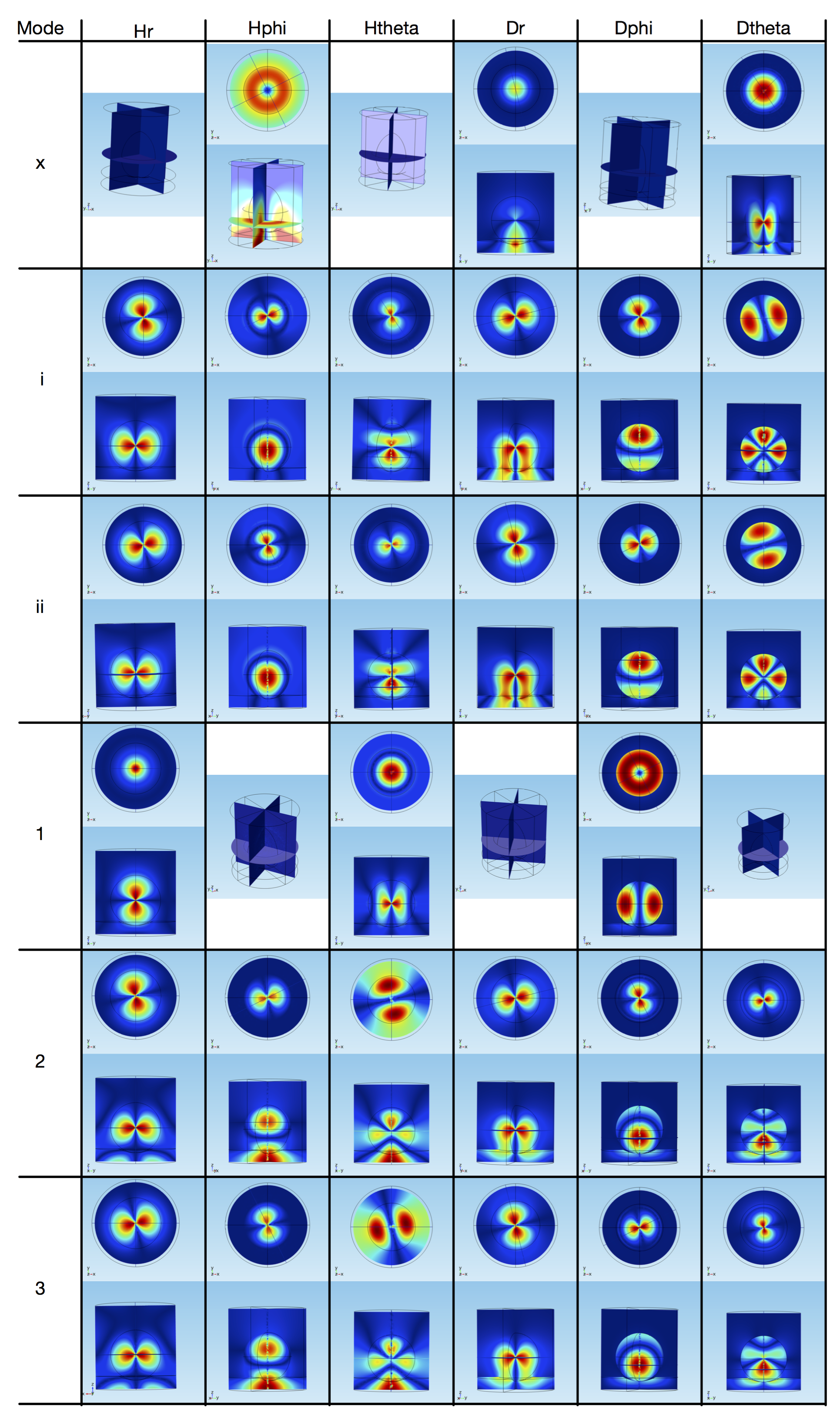}
\caption{Spherical coordinate field components of the six lowest dielectric modes in the YIG/sapphire/air system. Each mode is viewed parallel to the $z$-axis (top row) and in the $x,y$ plane (bottom row), except where no field is present. We can readily identify modes $x$ and 1 as (0,0) and (1,0) spherical dielectric modes, respectively. These two modes appear as {``}pure{''} dielectric modes containing only three field components. The two doublet modes ($i$, $ii$, 2 and 3) appear to contain energy density in all six field components, and are greatly affected by the sapphire support, which is what appears to lead to the additional degeneracy; splitting two modes in to four modes. From the radial components we can identify these modes as being modified (1,1) spherical dielectric modes.
}
\label{fig:shapes}
\end{figure*}

The results of the FEM using a value of $\epsilon_\text{YIG}/\epsilon_0=15.965$ and $r^\prime=3.71$ mm, where $r^\prime$ is the radius of curvature of the sapphire support{'}s concavity, are summarised in figure \ref{fig:shapes} and in table \ref{tab:freqs2}. The measured frequency of the doublet modes has been taken as the average of the two constituent{'}s frequencies at $B=7$ T.

	\begin{table}[b!]
	\centering
	\begin{tabular}{|c|c|c|c|}
	\hline
	{~}Mode{~} & $f_\text{meas}$ (GHz)&$f_\text{sim}$ (GHz)&$(n,m)$\\
	\hline
	$x$ & 12.779 & 12.785 & (0,0)\\
	\hline
	$i$ $\&$ $ii$ & 15.534 & 15.286 & (1,1)\\
	\hline
	1 & 15.732 & 15.736 & (1,0)\\
	\hline
	2 $\&$ 3 & 15.922 & 15.921 & (1,1)\\
	\hline
	\end{tabular}
	\caption{Comparison of FEM and measured frequencies and hence mode identification.}
	\label{tab:freqs2}
	\end{table}
From the FEM and the analytical mode shapes of spherical dielectric resonances described by \cite{spheres}, we can identify mode $x$ as an $n=0$ mode with no degeneracy. Therefore it is present as a singular resonance. The other five modes appear as $n=1$ modes. There should only exist a $2n+1$ fold degeneracy for resonant spherical photon modes, which can be broken by internal impurities or by asymmetric boundary conditions set by a cylindrical enclosure, microwave loop probes, and the sapphire substrate, collectively termed {``}backscatterers{''}. This degeneracy arises from a Legendre polynomial in the mode{'}s $H$ and $E$ field analytical expressions of the form $P^m_n(\cos\theta)\left\{^{\cos(m\phi)}_{\sin(m\phi)}\right\}$, where $m=0, \ldots , n$. The integers $m$ and $n$ represent the number of maxima of the mode{'}s energy density in the $\phi$ direction over 180$^\circ$, and the number in the $\theta$ direction over 180$^\circ$, respectively. This would imply that for $n=1$ we should observe three distinct modes corresponding to a single $(n,m)=(1,0)$ and two $(1,1)$ modes, rather than five modes. However, the FEM demonstrates that the use of the sapphire support base introduces a further degeneracy to the $(1,1)$ modes depending on the amount of field that permeates the sapphire. The modelling predicts four $(1,1)$ modes, existing as two sets of two, which are separated by approximately 500 MHz. This is in fair agreement with the separation of modes $i$, $ii$ with modes 1{--}3. Therefore it is apparent that modes $i$ and $ii$ are a doublet pair with $(n,m)=(1,1)$. 

Given that modes 2 and 3 approach relatively similar frequencies at high magnetic fields, it is reasonable to assume that these modes correspond to the second $(1,1)$ doublet pair, which FEM predicts will have a larger proportion of microwave field inside the sapphire support. This means that mode 1 must be the $(1,0)$ single mode. 

%given that it is offset from the lower and upper modes by $\sim$150 MHz and $\sim$ 200 MHz, values too large to originate from the typical size of mode{--}mode couplings that break the degeneracy between electromagnetic resonances in the absence of magnons. 

From figure \ref{fig:modefits}, we can see that both the doublet pairs demonstrate a gyrotropic response when interacting with the magnon resonances, i.e., one mode interacts more than its doublet pair. This is a common occurrence in spin ensemble systems and has been observed in paramagnetic systems such as Fe$^{3+}$ in sapphire \cite{gyro1,gyro2,gyro3}. This asymmetric interaction strength for doublet pairs has also been observed in ferromagnets by Krupka \textit{et al.} \cite{krupka1,krupka2} and predicted by Rameshti \textit{et al.} \cite{Rameshti}, with the latter stating that $g_{n,m=n}>g_{n,m=-n}$, where a different notation to that used here is employed, in which $m=-n,\ldots,0,\ldots,n$. The notations are equivalent as an $m=\pm n$ doublet in \cite{Rameshti} corresponds to a $\left\{^{\cos(m\phi)}_{\sin(m\phi)}\right\}$ doublet pair here.

The gyrotropic response is a result of the anisotropy of a ferromagnet{'}s permeability tensor; the same reason why these materials are used in circulators. The permeability tensor, containing off diagonal terms appears as
\begin{equation}
\vec{\mu}=\mu_0\left(\begin{matrix}
1+\chi & -i \kappa & 0 \\
i\kappa & 1+\chi & 0 \\
0 & 0 & 1 
\end{matrix}\right),
\label{eq:mu}
\end{equation}
where $\mu_0$ is the permeability of free space, and $\chi$ is the magnetic susceptibility of the ferromagnet, which is related to the magnetic permeability tensor by $\vec{\mu}=\mu_0\left(\vec{1}+\vec{\chi}\right)$.

When any resonant photonic mode exists as a doublet, it is because the  $\left\{^{\cos(m\phi)}_{\sin(m\phi)}\right\}$ degeneracy has been broken by some backscatterer, and the two resulting modes exist as counter propagating travelling waves \cite{gyro1,gyro2}. The overall effect is that one travelling wave will see an effective permeability of $\mu_+=\mu_0(1+\chi+\kappa)$, whilst the other will see $\mu_-=\mu_0(1+\chi-\kappa)$, which can be rewritten as $\mu_\pm=\mu_0(1+\chi_\pm)$, and we can state that $(\chi_++\chi_-)/2=\chi$, where $\chi$ is the {``}unperturbed{''} magnetic susceptibility that a standing wave would observe.

 \begin{table}[b!]
\centering
\begin{tabular}{|c|c|c|c|c|}
\hline 
{~}Mode{~} & {~}$\omega_{j\left| B\to\text{7 T}\right.}/2\pi$ & {~}$g_j/\pi${~} & $\xi_j$ & $\chi_\text{eff}$\\
&(GHz)&(GHz)&&\\ \hline
$x$ & 12.779 & 4.79 & 0.221 & 0.159\\ \hline
$i$ & 15.506 & 7.11 & 0.594 & 0.0885\\ \hline
$ii$ & 15.563 & 4.19 & 0.594 & 0.0305\\ \hline
1& 15.732 & 6.15 & 0.728 & 0.0525\\ \hline
2 & 15.893 & 3.04 & 0.493 & 0.0185\\ \hline
3 & 15.950 & 0.78 &{~} {~}0.493{~}{~} & {~}{~}0.00121{~}{~}\\ \hline
\end{tabular}
\caption{Calculated magnetic filling factors, $\xi_j$ and effective magnetic susceptibilities, $\chi_\text{eff}$ for each of the photon modes.}
\label{tab:ff}
\end{table}

The effective susceptibility that a mode experiences will determine the interaction strength of that mode with a magnon resonance according to \cite{Goryachev}:
\begin{equation}
g_i^2=\chi_\text{eff}\omega^2\xi,
\end{equation}
where $\xi$ is the total magnetic filling factor of the mode; i.e., the proportion of magnetic field within the ferromagnetic material compared to the entire system. This parameter is used in an attempt to quantify the overlap of the magnon and photon modes and is calculated as
	\begin{equation}
	\xi=\frac{\int\int\int_{V_\text{YIG}} \mu_0 \vec{H}^*\vec{H} dV_\text{YIG}}{\int\int\int_V \mu_0\vec{H}^*\vec{H}dV}{~}.
	\label{eq:ff}
	\end{equation}
It should be noted that typically it is only the magnetic field energy density perpendicular to the external magnetic field that is considered to interact with the spin system \cite{Goryachev,Soykal}. However, the interaction of mode 1 is far larger than its perpendicular filling factor of 0.075 would suggest. So, in an attempt to account for the interaction with nonuniform magnon modes, the total magnetic filling factor has been used. These have been calculated from the FEM and the resulting values of $\chi_\text{eff}$ are displayed in table \ref{tab:ff}.

Given our assumption that mode 1 represents the (1,0) dielectric mode, which will exist as a standing wave given no possible degeneracy, the calculated $\chi_\text{eff}$ value for this mode should represent the unperturbed magnetic susceptibility of the YIG. Taking the average of the $\chi_\text{eff}$ values for the doublet modes $i$ ($\chi_+$) and $ii$ ($\chi_-$) yields a value of $\chi=0.0595$; in reasonable agreement with the value obtained from mode 1. 

The FEM predicts that modes $x$, 2 and 3 will each contain a significant proportion of magnetic field energy within the sapphire support, so one would expect these modes to observe a lower effective magnetic susceptibility, which would appear true for the latter two modes (their average susceptibility yields an unperturbed susceptibility of $\sim0.01$). However, mode $x$ demonstrates a much larger coupling strength than what should be afforded a mode with its filling factor, hence a $\chi_\text{eff}$ value approximately three times larger than the unperturbed value obtained from modes $i$, $ii$ and 1. This suggests that our approximation of using the total magnetic filling factor to quantify the overlap of the magnon and photon modes is not entirely accurate. To accurately explain the origins of the differing interaction strengths of each mode, knowledge of higher order, nonuniform magnon mode shapes are required, in order to replace the filling factor approximation with an overlap value. Unlike Zhang \textit{et al.}{'}s \cite{Zhang} ultrastrong coupling results with a $d=2.5$ mm YIG sphere, in which higher order magnon modes mostly couple weakly with the microwave cavity, here we excite internal, nonuniform electromagnetic resonances, so it is more likely than not that these modes will couple more strongly to nonuniform magnon modes if their mode shapes match up well spatially. The derived values of susceptibility in table \ref{tab:ff} agree within an order of magnitude to previously measured results \cite{krupka1} but have been underestimated due to the use of filling factor as opposed to a mode overlap integral.

\begin{figure}[t!]
\centering
\includegraphics[width=0.45\textwidth]{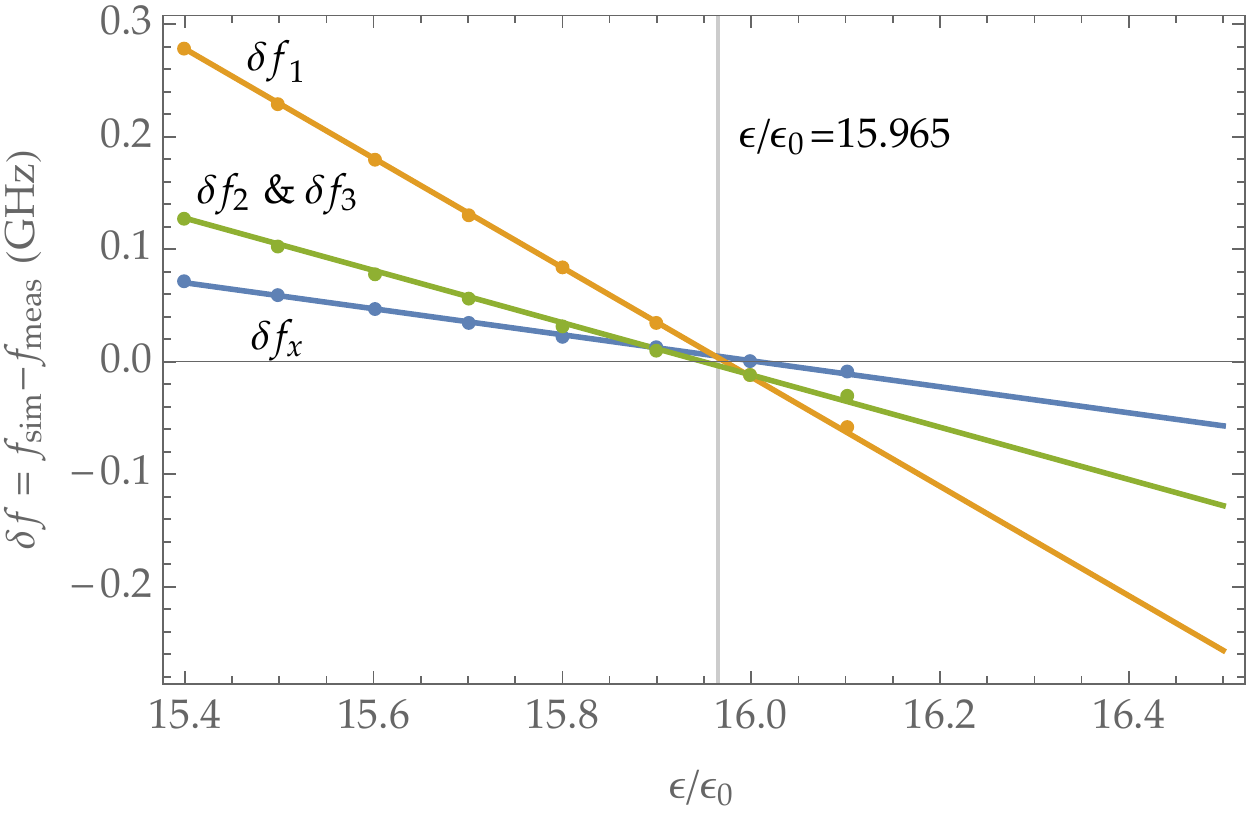}
\caption{Frequency difference between simulated and measured results as the relative permittivity of YIG is varied in the FEM software. The radius of curvature of the sapphire support used here was $r^\prime=3.7$ mm, which for mode 1 is largely irrelevant, but for modes $x$, 2 and 3 gives good agreement. }
\label{fig:perm}
\end{figure}

Finally, we can use the predicted mode frequencies of the FEM to determine the permittivity of the YIG sample, by varying $\epsilon_\text{YIG}/\epsilon_0$ until the frequencies match the asymptotic values measured at high magnetic fields. At these magnetic field values, the matrix in equation (\ref{eq:mu}) becomes the identity matrix \cite{krupka1}. By measuring the depth of the sapphire concavity and its width at the surface, the radius of curvature was determined to be $r^\prime=3.71\pm0.2$ mm. With this information, an iterative simulation was conducted mapping mode frequencies versus relative permittivity of YIG. It was found that mode 1 is relatively insensitive to the radius of curvature of the sapphire support. This is due to the absence of electric field density outside the YIG for this particular mode. Given that $r^\prime$ contains a significant amount of uncertainty, this mode is used to match $f_\text{sim}$ with $f_\text{meas}$. A plot of $\delta f=f_\text{sim}-f_\text{meas}$ versus permittivity is shown in figure \ref{fig:perm}. From this result, we can state that $\epsilon_\text{YIG}/\epsilon_0=15.96\pm 0.02$. This value agrees well with previous measurements taken using the so-called {``}Courtney{''} technique with YIG samples \cite{krupka2}.\\

In conclusion, we observe ultrastrong coupling between internal dielectric microwave resonances and magnons inside a $d=5$ mm YIG sphere. The large diameter of the sphere results in not only an increased number of spins, but also the accessibility of the internal electromagnetic resonances due to their existence below K-band frequencies. The use of internal microwave modes instead of an external cavity resonance results in far larger magnetic filling factors than ever before achieved in such an experiment, hence the coupling values and cooperitivity values observed are, to the authour{'}s knowledge, the largest ever reported, with a maximum $g/\pi=7.11$ GHz, or $\sim7000$ mode linewidths, and $C=1.5\times10^7$. This implies an extremely high level of coherence in this system. Most importantly however, the numerous resonant magnon peaks in the dispersive regime and the discrepancies in calculated susceptibilities suggest that higher order magnon modes participate in this system. This implies that the previously theoretically analysed models of such systems are incomplete. 

\bibliography{YIGspheres.bib}

\end{document}